\documentclass[twocolumn,aps,prb,floatfix,showpacs]{revtex4}
\usepackage{bm}
\usepackage{amssymb,amsmath}
\usepackage{dcolumn}
\usepackage{graphicx}
\usepackage{bbold}
\usepackage{booktabs}
\usepackage{color}
\usepackage[colorlinks,bookmarks=false,citecolor=blue,linkcolor=blue,urlcolor=blue]{hyperref}

\def\nhat{\hat{n}}
\def\ahat{\hat{a}}

\def\fhat{\hat{f}}

\def\Xhat{\hat{X}}
\def\calXhat{\hat{\cal X}}
\def\That{\hat{T}}
\def\calThat{\hat{\mathcal{T}}}
\def\Nhat{\hat{N}}
\def\Mhat{\hat{M}}
\def\alphahat{\hat{\alpha}}

\def\phihat{\hat{\phi}}
\def\psihat{\hat{\psi}}
\def\r{{\bf r}}

\def\q{{\bf q}}

\def\calU{{\cal U}}

\def\S{{\bf S}}

\begin{document}
\title{Exact solution of the infinite-$U$ Hubbard problem and other models in one dimension}
\author{Brijesh Kumar}
\email{bkumar@mail.jnu.ac.in}
\affiliation{School of Physical Sciences, Jawaharlal Nehru University, New Delhi 110067, India.}
\date{\today}

\begin{abstract}
An interacting spin-fermion model is exactly solved on an open chain. In a certain representation, it is the nearest-neighbor Hubbard model in the limit of infinite $U$ (local interaction). Exact solution of its complete energy eigen-spectrum is accomplished by introducing a unitary transformation which maps the original problem to a tight-binding model of the fermions only. Physically,  the exact solution implies the {\em absence} of Nagaoka ferromagnetism in the ground state for arbitrary electron densities. The present method solves a class of very general models exactly. Few more problems are discussed as an application of this unitary transform method.
\end{abstract}

\maketitle
\section{\label{sec:intro} Introduction}
The physics of strongly correlated lattice electrons is complex as well as interesting. The electronic properties of very many exciting materials (such as, the transition metal oxides, rare earths etc.) exhibit  this physics in a variety of different ways. A minimal model for understanding these systems requires one to consider, at least, the local coulomb repulsion, in addition to the tight-binding hopping of electrons. In an effective one-band system, the local repulsion, $U$, is the energy cost for putting two electrons with opposite spins together on the same site (Wannier orbital). This model, famously called the Hubbard model, has been a subject of great interest. It poses one of the most difficult problems in quantum many body theory. Although the model has been long solved exactly in one dimension (1D) using Bethe ansatz~\cite{Lieb_Wu}, and also shown to be integrable~\cite{Shastry}, it still evades an exact analytical solution in higher dimensions. In recent times, the interest in the Hubbard model has been further reinforced by the high-T$_C$ superconductivity in cuprates, and also by the need to understand other phenomena in strongly correlated electrons~\cite{Imada}.

There has been a long-standing interest in understanding the metallic-ferromagnetism through Hubbard model. The limit of infinite local repulsion in one-band Hubbard model presents an interesting case study in this context. In this limit, the ground state on certain lattices (for example, square lattice) was shown to be saturated metallic-ferromagnetic for a single hole in an otherwise half-filled system (Nagaoka-Thouless theorem)~\cite{Nagaoka,Thouless}. Subsequent variational and numerical studies have shown that the Nagaoka ferromagnetism survives for finite hole densities away from half-filling, up to a (lattice dependent) critical doping~\cite{Nagaoka_SKA, Linden, Becca, DMFT}. The Lieb-Mattis theorem, however, rules out the existence of ferromagnetism in 1D~\cite{Lieb_Mattis_Theorem}. For the infinite-$U$ Hubbard model with nearest-neighbor hopping, it implies the lack of Nagaoka ferromagnetism, which is borne out by the analytic studies of this problem for one and two holes explicitly. These investigations either use the Bethe ansatz approach~\cite{Doucot_Wen}, or work with an effective spin Hamiltonian in the presence of a single hole~\cite{Haerter_Shastry}, starting with a Gutzwiller projected Hamiltonian.

Recently, we have developed a new approach to the infinite-$U$ Hubbard problem~\cite{BK_new_rep}. In our formulation, we canonically represent an electron in terms of a spinless fermion and the spin-1/2 (Pauli) operators. We then write the Hubbard model in this representation. Finally, we take the limit $U\rightarrow\infty$, and get the infinite-$U$ Hamiltonian, $H_\infty$. Although our prescription is applicable to any lattice, it clearly distinguishes the bipartite lattices from others. By exploiting the two sublattice structure of a bipartite lattice, we can write $H_\infty$ in a beautiful form (resembling the Anderson-Hasegawa Hamiltonian, but different from it and fully quantum mechanical) which reveals the phenomenon of metallic-ferromagnetism in the infinite-$U$ Hubbard model in a very transparent way. (Otherwise, we all know that the Nagaoka ferromagnetism is a non-obvious strong correlation effect.)

In the present study, we investigate 1D infinite-$U$ Hubbard problem within our approach, and exactly solve it by means of the suitably constructed unitary transformation. From this exact solution, we conclude that its ground state is `correlated metallic' and 'ideal paramagnetic' for arbitrary density of electrons. Ours is a `non-Bethe' method which also solves a class of very general models in 1D. The infinite-$U$ Hubbard model happens to be one among them. This paper is organized as follows. First we present the exact solution of $H_\infty$ in 1D. Then, we identify a general class of models which can be exactly solved by our method. We also briefly discuss the Anderson-Hasegawa problem and the minimal coupling lattice Hamiltonian in the light of ideas developed here. Finally, we conclude with a summary.
\section{\label{sec:1D_Hubbard} Infinite-$U$ Hubbard model in 1D}
According to a recently developed canonical (and invertible) representation, an electron can be described in terms of a spinless fermion and the Pauli operators~\cite{BK_new_rep}. On a bipartite lattice, we can represent the electrons on different sublattices in two different, but equivalent, ways. (In principle, we can generate infinitely many equivalent representations through unitary transformations, but our purpose is served by the following two forms.)
The electronic operators in this representation are: $ \fhat^\dag_{l,\uparrow}=\phihat^{ }_l\sigma^+_l$, $\fhat^\dag_{l,\downarrow}=(i\psihat_l-\phihat_l\sigma^z_l)/2$ 
on the odd-numbered sites ($l=1,3,5,\dots$), and $\fhat^\dag_{l,\uparrow}=i\psihat_l\sigma^+_l$, $\fhat^\dag_{l,\downarrow}=(\phihat_l-i\psihat_l\sigma^z_l)/2$ on the even-numberd sites, where  $\phihat_l=(\ahat^\dag_l+\ahat_l)$ and $i\psihat=(\ahat^\dag_l-\ahat_l)$ are the Majorana fermions. Here, $\ahat^{ }_l$ and $\ahat^\dag_l$ are the spinless fermion operators, and $\vec{\sigma}_l$ the Pauli operators, on the $l^{th}$ site.
Moreover, the electronic number operator is written as: $\Nhat_e = L+\sum_{l=1}^L(1-\nhat_l)\sigma^z_l$, and the `physical spin' of an electron on the $l^{th}$ site is given by: $\vec{\S}_l = \nhat_l\vec{\sigma}_l/2$, where $\nhat_l = \ahat^\dag_l\ahat^{ }_l$. For completeness, also note the local mapping: $\left|0\rangle\right.=\left|\stackrel{-}{_\circ}\right\rangle$,  $\left|\uparrow\rangle\right.=\left|\stackrel{+}{_\bullet}\right\rangle$, $\left|\downarrow\rangle\right.=\left|\stackrel{-}{_\bullet}\right\rangle$, and $\left|\uparrow\downarrow\rangle\right.=\left|\stackrel{+}{_\circ}\right\rangle$ (on an odd-numbered site) and $=-\left|\stackrel{+}{_\circ}\right\rangle$ (on an even-numbered site). Here, $\left|0\rangle\right.$, $\left|\uparrow\rangle\right.$, $\left|\downarrow\rangle\right.$ and $\left|\uparrow\downarrow\rangle\right.$ are the local electronic states (in the usual notation). The states $|\circ\rangle$ and $|\bullet\rangle$ denote the empty and filled site, respectively, for the spinless fermion, and $\{|+\rangle,|-\rangle\}$ is the basis set of a single Pauli spin. Note that $|+\rangle$ and $|-\rangle$ represent the actual electronic spin on a site when it is occupied by a spinless fermion.

Now, consider the Hubbard model with nearest-neighbor hopping on an open chain. (The reason for working with an `open' chain will become clear shortly.) Since the hopping process is bipartite, we can use the above two forms of the representation for electrons to convert the Hubbard model into a corresponding `spin-fermion' model. In the limit of infinite-$U$, we get the following Hamiltonian. 
\begin{equation}
H_\infty = -t\sum_{l=1}^{L-1}\Xhat^{ }_{l,l+1}\left(\ahat^\dag_{l}\ahat^{ }_{l+1} + \ahat^\dag_{l+1}\ahat^{ }_l\right) \label{eq:H_infty}
\end{equation}
For the details of its derivation, please refer to Ref.~[11].
Here, $\Xhat^{ }_{l,l+1}=(1+\vec{\sigma}_l\cdot\vec{\sigma}_{l+1})/2$ is the Dirac-Heisenberg exchange operator, and $L$ is the total number of lattice sites.  

\subsection{\label{subsec:exact_sol} Exact analytic solution}
For the moment, we forget the physical origin and purpose of $H_\infty$, and just take it as a given spin-fermion model in one dimension. Our immediate goal is then to find its eigenvalues and eigenstates. There are two key features that we make use of in exactly solving this problem. First is the property, $\Xhat^{2}_{l,l+1}=\mathbb{1}$, of the exchange operators. And, the second is the open boundary condition of the 1D lattice (similar to the $XY$ spin-1/2 chain~\cite{XYchain}). 
We exploit the former to construct a unitary operator which, on an open chain, transforms Eq.~(\ref{eq:H_infty}) to a tight-binding model of the spinless fermions only. 

We first define the following unitary operator on the first bond [that is, for the pair of sites (1,2)].
\begin{equation}
\calU_{1,2}=\left(1-\nhat_2\right) + \nhat_2 \Xhat_{2,1} \label{eq:U12}
\end{equation}
Here, $\Xhat^{ }_{2,1}=\Xhat^{ }_{1,2}$. Clearly, $\calU^\dag_{1,2}=\calU^{ }_{1,2}$ and $\calU^2_{1,2}=\mathbb{1}$. Thus, $\calU_{1,2}$ is both Hermitian as well as unitary. Moreover, it has the following important property.
\begin{equation}
\calU^\dag_{1,2}\left( \ahat^\dag_1 \Xhat^{ }_{1,2} \ahat^{ }_2\right)\calU^{ }_{1,2} = \ahat^\dag_1\ahat^{ }_2
\end{equation}
In the above equation, $\calU_{1,2}$ leaves $\ahat_1$ and $\Xhat_{1,2}$ unaffected while transforming $\ahat_2\rightarrow \Xhat_{2,1}\ahat_2$. Thus, $\calU_{1,2}$ gets rid of the exchange operator  $\Xhat_{1,2}$, and what remains is the hopping of the fermions alone. As it happens, we will get rid of the exchange operators on each bond by carefully following this approach. 

Before constructing a similar $\calU_{2,3}$ for the next bond, it is important to consider the effect of $\calU_{1,2}$ on other terms in $H_\infty$. Clearly, $\calU_{1,2}$ leaves the operators on other bonds unaffected, except for the bond (2,3). The net effect of this unitary transformation on $H_\infty$ is the following.
\begin{widetext}
\begin{equation}
\calU_{1,2}^\dag \, H_\infty \, \calU_{1,2} = -t\left\{\left(\ahat^\dag_1\ahat^{ }_2 + \ahat^\dag_2\ahat^{ }_1 \right) + \left(\ahat^\dag_2 \, \Xhat_{1,2}\Xhat_{2,3} \, \ahat^{ }_3 + \ahat^\dag_3 \, \Xhat_{3,2}\Xhat_{2,1} \, \ahat^{ }_2\right) + \sum_{l=3}^{L-1}\left(\ahat^\dag_l \, \Xhat^{ }_{l,l+1} \, \ahat^{ }_{l+1} + \ahat^\dag_{l+1} \, \Xhat^{ }_{l+1,l} \, \ahat^{ }_l \right)\right\}
\label{eq:H_12}
\end{equation}
\end{widetext}
The $\calU_{1,2}$ transfers $\Xhat_{1,2}$ from the first bond to the second. 
We define $\calXhat_{2,3}=\Xhat_{1,2}\Xhat_{2,3}$ and $\calXhat_{3,2}=\Xhat_{3,2}\Xhat_{2,1}$, which replace $\Xhat_{2,3}$ on bond (2,3). Note that $\calXhat^\dag_{2,3}=\calXhat_{3,2}\neq\calXhat_{2,3}$ and $\calXhat^\dag_{2,3}\calXhat^{ }_{2,3}=\calXhat^{ }_{3,2}\calXhat^{ }_{2,3}=\mathbb{1}$. Thus, $\calXhat_{2,3}$ is unitary but not Hermitian (unlike $\Xhat_{2,3}$).

Now we define the following unitary operator for getting rid of $\calXhat_{2,3}$ and $\calXhat_{3,2}$  from the terms inside the second parentheses in Eq.~(\ref{eq:H_12}).
\begin{equation}
\calU_{2,3}=\left(1-\nhat_3\right) + \nhat_3\calXhat_{3,2} \label{eq:U23}
\end{equation}
Note that $\calU_{2,3}$ is unitary but not Hermitian (unlike $\calU_{1,2}$). We can show that $\calXhat_{2,3}$ is invariant under $\calU_{2,3}$, while
\begin{equation}
\calU_{2,3}^\dag \, \ahat^{ }_3 \, \calU^{ }_{2,3} = \ahat^{ }_3 \, \calXhat^{ }_{3,2}.
\end{equation}
Therefore,
\begin{equation}
\calU^\dag_{2,3}\left( \ahat^\dag_2 \, \calXhat^{ }_{2,3} \, \ahat^{ }_3\right)\calU^{ }_{2,3} = \ahat^\dag_2\ahat^{ }_3.
\end{equation}
Moreover, $\calU^\dag_{2,3}\,(\ahat^\dag_{3} \Xhat_{3,4} \ahat^{ }_4)\, \calU^{ }_{2,3} = \ahat^\dag_{3}\calXhat_{3,4}\ahat^{ }_4$, where $\calXhat_{3,4}=\calXhat_{2,3}\Xhat_{3,4}=\Xhat_{1,2}\Xhat_{2,3}\Xhat_{3,4}$. It is clear by now that we can continue this process, and get rid of all the exchange operators in $H_\infty$. To achieve this, we define the following general unitary operators,
\begin{eqnarray}
\calU_{l,l+1} &=& \left(1-\nhat_{l+1}\right) + \nhat_{l+1}\calXhat_{l+1,l} \\
\calU &=& \prod_{l=1}^{L-1}\calU_{l,l+1} \label{eq:full_U}
\end{eqnarray}
where $\calXhat_{l,l+1} = \calXhat_{l-1,l}\Xhat_{l,l+1}=\prod_{m=1}^{l}\Xhat_{m,m+1}$, for $l=2, L-1$. We can show that the full unitary operator $\calU$, of Eq.~(\ref{eq:full_U}), transforms $H_\infty$ to a Hamiltonian of free spinless fermions. That is, 
\begin{eqnarray}
\calU^\dag \, H_\infty \, \calU &=& -t\sum_{l=1}^{L-1}\left(\ahat^\dag_l\ahat^{ }_{l+1} + \ahat^\dag_{l+1}\ahat^{ }_l\right) \label{eq:H_freefermi} \\
&=&\sum_k\epsilon_k~\ahat^\dag_k\ahat^{ }_k \label{eq:H_diag}
\end{eqnarray}
where $\epsilon_k = -2t\cos{k}$, and $k$ is the momentum. The operators $\{\ahat_k\}$ are the fermions in the momentum space. Since Eq.~(\ref{eq:H_freefermi}) is derived on a chain with open boundary condition, the Fourier transformation between $\{\ahat_l\}$ and $\{\ahat_k\}$ is defined as: $\ahat_l= \sqrt{\frac{2}{L+1}}\sum_k \ahat_k\, \sin{kl}$, where $k=n\pi/(L+1)$ for $n=1,2,\cdots, L$. Hence, the exact solution of $H_\infty$ in 1D. 
To this end, we also note that 
\begin{equation}
\calU^\dag\,\Nhat\,\calU = \Nhat ~~\mbox{and}~~ \calU^\dag\,\Mhat_\sigma \,\calU = \Mhat_\sigma,
\label{eq:N_and_Msigma}
\end{equation}
where $\Nhat = \sum_{l=1}^L\nhat_l$ is the number operator of the spinless fermions, and $\Mhat_\sigma = \sum_{l=1}^L\sigma^z_l$ is the total $\sigma^z$ operator. Therefore, $\calU$ diagonalizes a more general Hamiltonian: $H=H_\infty -\lambda\Nhat -\eta\Mhat_\sigma$, where $\lambda$ is the chemical potential of the spinless fermions and $\eta$ is the external `magnetic' field acting on  the spins.

The above exercise presents a rigorous and transparent case of the complete decoupling of the Pauli and Fermi attributes of the electron. 
As a result of this decoupling, the energy eigenvalues of $H_\infty$ become independent of spins, giving rise to an extensive entropy in every eigenstate. Just as the spinless fermions $\{\ahat_l\}$ transform under $\calU$, the spins $\{\vec{\sigma}_l\}$ also transform to new spins. However, the total $\sigma^z$ is invariant under $\calU$, as noted in Eq.~(\ref{eq:N_and_Msigma}). It is these transformed spins that  are absent in the free spinless fermion Hamiltonian [Eq.~(\ref{eq:H_freefermi})].

\subsection{\label{subsec:no_Nagaoka} Absence of the Nagaoka ferromagnetism in 1D} 
In order to discuss the ground state of the infinite-$U$ Hubbard model, first let us recall the complete connection between the problem worked out in the previous subsection and the physical Hubbard model. Technically, the infinite-$U$ Hubbard model in our representation is not just $H_\infty$, but $H_\infty + \infty\sum_l(\frac{1}{2}-\nhat_l)$. That is, the physical problem is described by $H_\infty$ with an infinite chemical potential for the spinless fermions (please look into Ref.~[11] for details). Next, we note that $N=N_e-2N_D$, where $N$ is the total number of spinless fermions (that is, the number of singly occupied sites in terms of  the electrons), $N_e$ is the total number of electrons, 
and $N_D$ is the total number of doubly occupied sites.  Clearly, $N$ is a conserved quantity of $H_\infty$ [evident from Eq.~(\ref{eq:H_infty})]. So is $N_e$, and hence $N_D$. It is evident from the following explicit form of $H_\infty$ in terms of the electron operators.
\begin{equation}
H_\infty = -\frac{t}{2}\sum_{l=1}^{L-1}\sum_s^{\uparrow,\downarrow}\left(\nhat_{l,s}-\nhat_{l+1,s}\right)^2\left[\fhat^\dag_{l,\bar{s}}\fhat^{ }_{l+1,\bar{s}} + h.c \right]
\end{equation}

Since $N_e=N+2N_D$, we can label the sectors of states for a given $N_e$ in terms of the partitions: $(N,N_D)$. The physical validity of a partition, however, depends upon whether $N_e\le L$ (less than or equal to half-filling) or $N_e > L$ (more than half-filled case), subject to the natural constraint: $N+N_D \le L$. That is, an arbitrary partition of the integer $N_e$ into two other integers $N$ and $N_D$ does not necessarily denote a physical sector. For $0\le N_e\le L$,  the constraint is guaranteed to be satisfied. Therefore, all partitions are valid physical sectors. For example, if $N_e=7$ and $\le L$, then the corresponding sectors of states are: $(7,0)$, $(5,1)$, $(3,2)$, and $(1,3)$. However, when $N_e>L$, the constraint will disqualify many partitions. For example, if $L=4$ and $N_e=7$, then the only physical sector is: $(1,3)$, as the other partitions such as $(3,2)$ don't respect the constraint. In general, the physical sectors for $N_e\le L$ are given by the set: $\{(N_e-2N_D,N_D),~ \forall ~ N_D=0, 1,\cdots ,[N_e/2]\}$, where $[N_e/2]=N_e/2$ for even values of $N_e$ and $=(N_e-1)/2$ for odd values. For $N_e>L$, we use the relations: $N[2L-N_e] = N[N_e]$, and $N_D[2L-N_e] = L-N_e+N_D[N_e]$, to find the physical sectors. These relations are a consequence of the particle-hole transformation on the electronic operators. Here, $N[N_e]$ and $N_D[N_e]$ denote the dependence of $N$ and $N_D$ on $N_e$. 

The exact solution of $H_\infty$ gives a highly disordered ground state in terms of the spinless fermions and the Pauli spins. However, we need to carefully translate its meaning for the electrons.  Interestingly, we are able to show that the ground state of the infinite-$U$ Hubbard model for a given $N_e$, is a fermi-sea which is $2^{N^*}$-fold degenerate [where $N^*$ is defined in Eq.~(\ref{eq:Nstar})]. Physically, it means that the ground state is metallic and ideally paramagnetic. In other words, it is not Nagaoka ferromagnetic. Nor it is a kinematic singlet (like a normal electronic fermi-sea).

Now, the {\em proof}. 
Due to the fact that $U=\infty$ and it is the chemical potential of the spinless fermions, the ground state of the Hubbard problem, for a given $N_e$, lies in the sector $(N^*,(N_e-N^*)/2)$, where $N^*$ is the maximum allowed value of $N$ for the given $N_e$. 
\begin{equation}
N^* = \left\{\begin{array}{lcl} N_e &,& 0\le N_e\le L\\ 2L-N_e &,& L\le N_e\le 2L\end{array}\right.
\label{eq:Nstar}
\end{equation}
Since $\Nhat$ is invariant under $\calU$ [Eq.~(\ref{eq:N_and_Msigma})], the infinite-$U$ ground state corresponds to the fermi-sea of  $N^*$ spinless fermions with dispersion $\epsilon_k$ [Eq.~(\ref{eq:H_diag})]. We derive the following exact expression for the ground state energy.
\begin{equation}
E_g[N^*] = -2t\cos{\left(\frac{\pi}{2}\frac{N^*+1}{L+1}\right)}\frac{\sin{\left(\frac{\pi}{2}\frac{N^*}{L+1}\right)}}{\sin{\left(\frac{\pi}{2}\frac{1}{L+1} \right)}}
\end{equation}
Expectedly, $E_g=0$ for $N_e=0$ and $2L$ (trivial cases: empty and fully filled bands, respectively), and also for $N_e=L$ (the half-filled case for $U=\infty$). In the thermodynamic limit ($L\rightarrow\infty$), for a finite electron density, $n_e=N_e/L$, the ground state energy density, $e_g=E_g/L$, can be written as~\cite{fnote_eg}:
\begin{equation} e_g[n_e]=-\frac{2t}{\pi}| \sin(\pi n_e)|. \label{eq:eg_ne} \end{equation}

Now, we enumerate the spin degeneracy in the ground state, which will decide the magnetic nature of the ground state. Let us discuss $N_e\le L$ case first. In this case, the ground sector is $(N_e,0)$. That is, the number of electrons is completely exhausted by the number of spinless fermions. As $N_D=0$, the remaining $L-N_e$ sites must be empty. The state of an empty site is uniquely $\left|\stackrel{-}{_\circ}\right\rangle$. However, the state of a site, occupied by a single electron, is either $\left|\stackrel{+}{_\bullet}\right\rangle$ or $\left|\stackrel{-}{_\bullet}\right\rangle$, corresponding to the fact that it could be an $\uparrow$ or $\downarrow$ spin electron. Therefore, corresponding to any given distribution of $N_e$ spinless fermions on $L$ sites (any one of the $^LC_{N_e}$ combinations), there are exactly $2^{N_e}$ states (a total of $^LC_{N_e}\times2^{N_e}$ states in the ground sector). For example, on a chain with $L=7$ and $N_e=3$, the states in the ground sector are like: $\left|\stackrel{\pm}{_\bullet}\stackrel{\pm}{_\bullet}\stackrel{\pm}{_\bullet}\stackrel{-}{_\circ}\stackrel{-}{_\circ}\stackrel{-}{_\circ}\stackrel{-}{_\circ}\right\rangle$, where the filled sites could be in any one of the $^7C_3$ combinations. On each filled site, the Pauli spin could be $+$ or $-$ (without affecting the number of electrons). Hence, $2^3$ different $M_\sigma$ states. 
Coming back to the general situation, these $2^{N_e}$ states can be grouped according to their $M_\sigma$ values ($^{N_e}C_{M_\sigma}$ states for a given $M_\sigma$). We know that $M_\sigma$ is a conserved quantity of $H_\infty$, and it is also invariant under $\calU$ [Eq.~(\ref{eq:N_and_Msigma})]. Therefore, $2^{N_e}$ different $M_\sigma$ states in the ground sector will be degenerate, as the exact energy eigenvalues of $H_\infty$ are independent of $M_\sigma$. Hence, the ground state, in the sector $(N_e,0)$, is a $2^{N_e}$-fold degenerate fermi-sea of $N_e$ spinless fermions. Since there are only empty or singly occupied sites, the $2^{N_e}$-fold degeneracy is strictly due to the physical spin of electrons.   Therefore, we conclude that the exact ground state of the infinite-$U$ Hubbard model is ideally paramagnetic and metallic (more correctly, {\em strange} or correlated metallic, as it is not a fermi-sea of the normal electrons). 

Furthermore, we note that the arguments for $N_e>L$ are the same as that for $N_e\le L$. The (only) key difference between the two is that the for $N_e>L$, there are only singly or doubly occupied sites in the ground sector, while for $N_e<L$ the sites are either empty or singly occupied. For example, when $N_e=11$ and $L=7$, then a typical state will be of the form: $\left|\stackrel{\pm}{_\bullet}\stackrel{\pm}{_\bullet}\stackrel{\pm}{_\bullet}\stackrel{+}{_\circ}\stackrel{+}{_\circ}\stackrel{+}{_\circ}\stackrel{+}{_\circ}\right\rangle$. This state is the counterpart of a previously mentioned state for less than half-filling. The physics of the Hubbard model on a bipartite lattice for $N_e$ electrons is same as that for $2L-N_e$ electrons. Without going into the (repetitive) details of the analysis all over again, we conclude that there is no Nagaoka ferromagnetism in the one-dimensional inifinte-$U$ Hubbard model, and that the exact ground state is a $2^{N^*}$-fold degenerate fermi-sea of $N^*$ spinless fermions, where $N^*$ is defined by Eq.~(\ref{eq:Nstar}).

To this end, we would like to make four comments. First is that the state labeling and counting procedure presented above is applicable to all lattices. It is not specific to 1D (although the solution is). Second comment is that, in the cases different from the present 1D problem, it will be impossible to completely get rid of the exchange operators. Due to which, the different $M_\sigma$ states are not guaranteed to be degenerate. Therefore, we stand a clear chance of finding some sort of metallic magnetism (ferro or antiferro~\cite{Haerter_Shastry} or something else) on other lattices (and hopping geometries). Third is a minor comment about $U=-\infty$ problem. In this case, the ground sector corresponds to $N_D=N_e/2$ (for even $N_e$) or $(N_e-1)/2$ (for odd $N_e$). The ground state (say for even $N_e$) is a  $^LC_{N_D}$-fold degenerate hard-core bosonic state with no kinetic energy gains (due to spinless fermion). The final comment is about the finite temperature calculations. Since $U$ is infinite, the ground sector (for a given $N_e$) is the only part of the Hilbert space which is accessible by finite temperatures. The thermodynamics of this problem can therefore be worked out in the canonical ensemble of the spinless fermions. The Pauli spins remain  ideally paramagnetic down to absolute zero temperature. This sets the quantum coherence temperature for electrons to be zero (even though there is this Fermi temperature scale for the spinless fermions)~\cite{TVR_comment}.

\subsection{\label{subsec:spin_orbit} Spin-orbital model}
In a different incarnation, the Hamiltonian $H_\infty$ can be considered as a one-dimensional model of the coupled spin and orbital degrees of freedom. By applying the Jordan-Wigner transformation on the spinless fermions, we can derive the following Hamiltonian,
\begin{equation}
H_{SO} = -t\sum_{l=1}^{L-1}\Xhat_{l,l+1}\left(\tau^+_l\tau^-_{l+1} + \tau^-_l\tau^+_{l+1}\right) - \frac{1}{2}\sum_l(\lambda\tau^z_l + \eta\sigma^z_l)
\label{eq:spin_orbit}
\end{equation}
where $\{\ahat_l\}$ have been changed to $\{\tau^-_l\}$, and $\tau^z_l = 2\nhat_l-1$. In a transition metal ion with two-fold orbital degeneracy, the orbital degrees of freedom can be described in terms of the Pauli operators. Let $\vec{\tau}$ denote this orbital degree in the present discussion. The model $H_{SO}$ is special case of the more general Kugel-Khomskii type models~\cite{Kugel_Khomskii}. Equation~(\ref{eq:spin_orbit}) thus  presents an exactly solvable spin-orbital lattice model, in which the spins and the orbitals behave as decoupled. While the orbital part acts as an $XY$ chain, the spins become paramagnetic.
\section{\label{sec:general_models} General class of exactly solvable models} 
While transforming $H_\infty$ to the tight-binding model of spinless fermions, it became clear that a very general class of models can be solved exactly by our method. A Hamiltonian in this class can be written as:
\begin{equation}
H = -t\sum_{l=1}^{L-1}\left(\ahat^\dag_{l} \That^{ }_{l,l+1}\ahat^{ }_{l+1} + \ahat^\dag_{l+1}\That^{ }_{l+1,l}\ahat^{ }_l\right) \label{eq:H_general}
\end{equation}
Here, $\That_{l,l+1}$ is some unitary operator on the bond $(l,l+1)$, and by defintion $\That_{l+1,l}=\That^\dag_{l,l+1}$. Moreover, $\That_{l,l+1}$ doesn't have to commute with $\That_{l-1,l}$  and $\That_{l+1,l+2}$, while strictly commuting with the $\That$ operators on other bonds, and also with the fermions, $\{\ahat_l\}$. With a few careful steps of algebra, this Hamiltonian can be transformed to a tight-binding model of the fermions [Eq.~(\ref{eq:H_freefermi})] with the help of $\calU = \prod_{l=1}^{L-1}\calU_{l,l+1}$, where $\calU_{l,l+1}=(1-\nhat_{l+1}) + \nhat_{l+1}\calThat_{l+1,l}$. Here, $\calThat_{l,l+1} = \prod_{m=1}^{l}\That_{m,m+1}$ and $\calThat_{l+1,l}=\calThat^\dag_{l,l+1}$.
For a special case in which $\That_{l,l+1}$ are just the phase factors $e^{i\xi_{l,l+1}}$, we can get rid of these on a Caley tree of arbitrary coordination, $z$ (for a nearest-neighbor chain, $z=2$). In this case, it can be done not only for the fermions but also for the bosons. 
\subsection{\label{subsec:spin-1} Spin-fermion model for higher spins}
As an academic exercise, we construct the spin-fermion models for higher spins, which belong to this general class of exactly solvable models. We achieve this by constructing the `correct' analog of the exchange operator for a pair of higher spins. By correct exchange we mean that $|m_1\rangle|m_2\rangle$ must become $|m_2\rangle |m_1\rangle$ under the exchange operator. For clarity, we work it out explicitly for the spin-1 case.  Here, $S^z|m\rangle =m|m\rangle$, with $m=1,0,\bar{1}$ (-1 is denoted as $\bar{1}$; for spin-$S$, $m=\bar{S}$, $\bar{S}+1$, $\dots$, $S$). The usual spin-spin interaction, $\S_1\cdot\S_2$, does not really exchange $m_1$ and $m_2$. Hence, we construct an operator $\Xhat_{1,2}$ such that $\Xhat_{1,2}|m_1\rangle|m_2\rangle = |m_2\rangle|m_1\rangle$. For example, $|0\rangle|0\rangle$ remains unaffected under $\Xhat_{1,2}$, while $|1\rangle|\bar{1}\rangle$ becomes $|\bar{1}\rangle|1\rangle$ and vice versa. Explicitly, in terms of the spin operators, this spin-1 exchange operator can be written as:
\begin{eqnarray}
\Xhat_{1,2} &=& 1-\left(S_{1z}^2 + S_{2z}^2\right) + \frac{1}{2}\S_1\cdot\S_2+\frac{1}{2}\left(\S_1\cdot\S_2\right)^2 + \nonumber\\
& &(\S_1\cdot\S_2)(S_{1z}S_{2z}) + \frac{i}{2}(\S_1\times\S_2)_z(S_{1z}-S_{2z}) \nonumber\\
& &-\frac{1}{2}[(\S_1\times\S_2)_z]^2 
\end{eqnarray}
where $(\S_1\times\S_2)_z = \frac{i}{2}(S_{1+}S_{2-}-S_{1-}S_{2+})$.
It is clear that we can similarly construct the exchange operators for higher spins. Since  $\Xhat^2_{1,2} = \mathbb{1}$, just like in the spin-1/2 problem, the corresponding spin-fermion model [that is, Eq.~(\ref{eq:H_infty})] can be diagonalized in the same way.
\subsection{\label{subsec:AH_Peierls} Some physical corrolaries}
\subsubsection{Anderson-Hasegawa problem}
Our method of getting rid of the unitary factors has an interesting consequence for the Anderson-Hasegawa (AH) problem. The AH model, $H_{AH}=-t\sum_{\langle l,m\rangle} \sqrt{\frac{1+{\bf \Omega}_l\cdot{\bf \Omega}_m}{2}}(\ahat^\dag_l e^{i\Phi_{l,m}}\ahat^{ }_m + h.c.)$, describes the motion of locally spin-projected electrons on a lattice, with  classical spins, $\{{\bf \Omega}_l\}$, in the background. Very often, the phases $\Phi_{l,m}$ arising due to the spin-projection along ${\bf \Omega}_l$ and ${\bf \Omega}_m$ are ignored without proper justification. There have been studies which rightly emphasize on taking into consideration the effects of these phases while computing physical properties~\cite{Pinaki}. The special case on a Caley tree discussed earlier, however, implies that indeed the original $H_{AF}$ can be transformed to $H_{AF}=-t\sum_{\langle l,m\rangle} \sqrt{\frac{1+{\bf \Omega}_l\cdot{\bf \Omega}_m}{2}}(\ahat^\dag_l \ahat^{ }_m + h.c.)$ on Caley trees. Thus, we give a reason for dropping the phases in $H_{AF}$, at least on a Caley tree. On an arbitrary lattice, however, one must keep these phases.
\subsubsection{Minimal coupling Hamiltonian in 1D}
We now briefly discuss the Peierls minimal coupling of the quantized electromagnetic radiation to the lattice fermions, in the light of these gauge removing tricks. Consider tight-binding electrons with nearest neighbor hopping on an open 1D lattice. The corresponding gauge-invariant Hamiltonian can be written as:
\begin{equation}
H=-t\sum_{l=1}^{L-1}\sum_{s}^{\uparrow,\downarrow}\left(\fhat^\dag_{l+1,s} \fhat^{ }_{l,s} e^{i\frac{e}{\hbar}\int_l^{l+1}A_x dx} + h.c.\right) + H_{F}
\label{eq:minimal}
\end{equation}
Here, $e=-|e|$ is the electronic charge, $H_F=\sum_{\q,\lambda}\omega_\q (\alphahat^\dag_{\q\lambda}\alphahat^{ }_{\q\lambda} +\frac{1}{2})$ is the field Hamiltonian (where $\alphahat_{\q\lambda}$, $\alphahat^\dag_{\q\lambda}$ are the Bose operators for an electromagnetic field of wavevector $\q$ and the polarization $\lambda$), and $A_x$ is the $x$-component of the vector potential. We have chosen the $x$-axis to be along the chain ($y=z=0$ line). In general, the vector potential can be written as:
\begin{equation}
\vec{A}(\r) = i\sum_{\q,\lambda} \sqrt{\frac{\hbar}{2\omega_\q\epsilon_0V}}\left[\vec{u}_{\q\lambda}(\r)\alphahat^\dag_{\q\lambda} - \vec{u}^*_{\q\lambda}(\r)\alphahat_{\q\lambda}\right]
\end{equation}
where $\vec{u}_{\q\lambda}$ is a normal mode (vector) function (including the information on polarization). Clearly, the vector potential operators at different spatial points commute with each other. The corresponding electric field operator is given by: $\vec{E}(\r)=\frac{i}{\hbar}[\vec{A}(\r),H_F]$.

Now again, we get rid of the field dependent factors from the hopping by applying the following unitary transformation:
\begin{equation}
\calU = \exp\left\{i\frac{e}{\hbar}\sum_{l=2}^L\nhat_l \int_{1}^l A_x dx\right\}
\end{equation}
where $\nhat_l = \nhat_{l,\uparrow} + \nhat_{l,\downarrow}$. Under this $\calU$, we get the following transformed Hamiltonian:
\begin{equation}
\calU^\dag H\,\calU = -t\sum_{l=1}^{L-1}\sum_s^{\uparrow,\downarrow}(\fhat_{l+1,s}^\dag\fhat^{ }_{l,s} + h.c.) + \calU^\dag H_F \calU
\end{equation}
While the hopping becomes simple, the field Hamiltonian transforms to:
\begin{eqnarray}
\calU^\dag H_F \calU &=& H_F - e\sum_{l=2}^L\nhat_l\int_1^l E_x dx + \nonumber \\ 
&& \frac{e^2}{2\epsilon_0}\sum_{l,l^\prime=2}^L V^{ }_{l,l^\prime} \nhat^{ }_l \nhat^{ }_{l^\prime}
\end{eqnarray}
Here, the second term is the potential energy in the presence of  electromagnetic field. Suppose $E_x$ is the electric field of a free radiation propagating in the $z$ direction. Then,  $e\sum_{l=2}^L\nhat_l\int_1^l E_x dx$ is same as $P_x E_x$, the dipole interaction. Here, $P_x=e\sum_{l=2}^L (l-1)\nhat_l$ is the electric polarization operator. Furthermore, $V_{l,l^\prime}=\frac{1}{V}\int_1^l dx\int_1^l dx^\prime \sum_{\q\lambda} \Re{[u^*_{x,\q\lambda}u_{x^\prime,\q\lambda}]}$ is the `Coulomb' repulsion between electrons, generated by the `exchange' of the photon (facilitated by $\calU$).  Thus, we have derived the gauge independent Coulomb and the dipole-field interactions, starting from the minimal coupling Hamiltonian on a lattice, without any approximations. Since the vector potential commutes at different points, in principle, we can do the same on a Caley tree as well.

\section{\label{sec:conclude} Conclusion}
To summarize, we have exactly solved the infinite-$U$ Hubbard model with nearest neighbor hopping on an open chain. We use a newly developed canonical representation for electrons, in which the Hubbard model becomes a spin-fermion model. This spin-fermion model is exactly solved by applying a   non-local unitary transformation. Under this transformation, the Pauli spins completely decouple from the  fermions, as a result of which, the ground state is correlated metallic and ideal paramagnetic for arbitrary density of electrons. 

This method solves a class of very general models. Guided by this observation, we have also constructed spin-fermion models for higher spins, by suitably extending the notion of `exchange' operators for higher spins. It is explicitly worked out for spin-1. (The spin only models, using our definition of the exchange operator for higher spins, exhibit interesting properties. These calculations will be discussed elsewhere.) By using the ideas developed here, we have shown that the phase factors in the Anderson-Hasegawa model can be droped on Caley trees. We have also derived the `dipole' interaction and Coulomb repulsion starting with Peierls minimal coupling Hamiltonian.
%
\bibliography{paper}

\begin{thebibliography}{19}
\expandafter\ifx\csname natexlab\endcsname\relax\def\natexlab#1{#1}\fi
\expandafter\ifx\csname bibnamefont\endcsname\relax
  \def\bibnamefont#1{#1}\fi
\expandafter\ifx\csname bibfnamefont\endcsname\relax
  \def\bibfnamefont#1{#1}\fi
\expandafter\ifx\csname citenamefont\endcsname\relax
  \def\citenamefont#1{#1}\fi
\expandafter\ifx\csname url\endcsname\relax
  \def\url#1{\texttt{#1}}\fi
\expandafter\ifx\csname urlprefix\endcsname\relax\def\urlprefix{URL }\fi
\providecommand{\bibinfo}[2]{#2}
\providecommand{\eprint}[2][]{\url{#2}}

\bibitem[{\citenamefont{Lieb and Wu}(1968)}]{Lieb_Wu}
\bibinfo{author}{\bibfnamefont{E.~H.} \bibnamefont{Lieb}} \bibnamefont{and}
  \bibinfo{author}{\bibfnamefont{F.~Y.} \bibnamefont{Wu}},
  \bibinfo{journal}{Phys. Rev. Lett.} \textbf{\bibinfo{volume}{20}},
  \bibinfo{pages}{1445} (\bibinfo{year}{1968}).

\bibitem[{\citenamefont{Shastry}(1986)}]{Shastry}
\bibinfo{author}{\bibfnamefont{B.~S.} \bibnamefont{Shastry}},
  \bibinfo{journal}{Phys. Rev. Lett.} \textbf{\bibinfo{volume}{56}},
  \bibinfo{pages}{2453} (\bibinfo{year}{1986}).

\bibitem[{\citenamefont{Imada et~al.}(1998)\citenamefont{Imada, Fujimori, and
  Tokura}}]{Imada}
\bibinfo{author}{\bibfnamefont{M.}~\bibnamefont{Imada}},
  \bibinfo{author}{\bibfnamefont{A.}~\bibnamefont{Fujimori}}, \bibnamefont{and}
  \bibinfo{author}{\bibfnamefont{Y.}~\bibnamefont{Tokura}},
  \bibinfo{journal}{Rev. Mod. Phys.} \textbf{\bibinfo{volume}{70}},
  \bibinfo{pages}{1039} (\bibinfo{year}{1998}).

\bibitem[{\citenamefont{Nagaoka}(1966)}]{Nagaoka}
\bibinfo{author}{\bibfnamefont{Y.}~\bibnamefont{Nagaoka}},
  \bibinfo{journal}{Phys. Rev.} \textbf{\bibinfo{volume}{147}},
  \bibinfo{pages}{392} (\bibinfo{year}{1966}).

\bibitem[{\citenamefont{Thouless}(1965)}]{Thouless}
\bibinfo{author}{\bibfnamefont{D.~J.} \bibnamefont{Thouless}},
  \bibinfo{journal}{Proc. Phys. Soc. (London)} \textbf{\bibinfo{volume}{86}},
  \bibinfo{pages}{893} (\bibinfo{year}{1965}).

\bibitem[{\citenamefont{Shastry et~al.}(1990)\citenamefont{Shastry,
  Krishnamurthy, and Anderson}}]{Nagaoka_SKA}
\bibinfo{author}{\bibfnamefont{B.~S.} \bibnamefont{Shastry}},
  \bibinfo{author}{\bibfnamefont{H.~R.} \bibnamefont{Krishnamurthy}},
  \bibnamefont{and} \bibinfo{author}{\bibfnamefont{P.~W.}
  \bibnamefont{Anderson}}, \bibinfo{journal}{Phys. Rev. B}
  \textbf{\bibinfo{volume}{41}}, \bibinfo{pages}{2375} (\bibinfo{year}{1990}).

\bibitem[{\citenamefont{Linden and Edwards}(1991)}]{Linden}
\bibinfo{author}{\bibfnamefont{W.}~\bibnamefont{Linden}} \bibnamefont{and}
  \bibinfo{author}{\bibfnamefont{D.~M.} \bibnamefont{Edwards}},
  \bibinfo{journal}{J. Phys.: Condens. Matter} \textbf{\bibinfo{volume}{3}},
  \bibinfo{pages}{4917} (\bibinfo{year}{1991}).

\bibitem[{\citenamefont{Becca and Sorella}(2001)}]{Becca}
\bibinfo{author}{\bibfnamefont{F.}~\bibnamefont{Becca}} \bibnamefont{and}
  \bibinfo{author}{\bibfnamefont{S.}~\bibnamefont{Sorella}},
  \bibinfo{journal}{Phys. Rev. Lett.} \textbf{\bibinfo{volume}{86}},
  \bibinfo{pages}{3396} (\bibinfo{year}{2001}).

\bibitem[{\citenamefont{Park et~al.}(2008)\citenamefont{Park, Haule,
  Marianetti, and Kotliar}}]{DMFT}
\bibinfo{author}{\bibfnamefont{H.}~\bibnamefont{Park}},
  \bibinfo{author}{\bibfnamefont{K.}~\bibnamefont{Haule}},
  \bibinfo{author}{\bibfnamefont{C.~A.} \bibnamefont{Marianetti}},
  \bibnamefont{and} \bibinfo{author}{\bibfnamefont{G.}~\bibnamefont{Kotliar}},
  \bibinfo{journal}{Phys. Rev. B} \textbf{\bibinfo{volume}{77}},
  \bibinfo{pages}{035107} (\bibinfo{year}{2008}).

\bibitem[{\citenamefont{Lieb and Mattis}(1962)}]{Lieb_Mattis_Theorem}
\bibinfo{author}{\bibfnamefont{E.~H.} \bibnamefont{Lieb}} \bibnamefont{and}
  \bibinfo{author}{\bibfnamefont{D.~C.} \bibnamefont{Mattis}},
  \bibinfo{journal}{Phys. Rev.} \textbf{\bibinfo{volume}{125}},
  \bibinfo{pages}{164} (\bibinfo{year}{1962}).

\bibitem[{\citenamefont{Doucot and Wen}(1989)}]{Doucot_Wen}
\bibinfo{author}{\bibfnamefont{B.}~\bibnamefont{Doucot}} \bibnamefont{and}
  \bibinfo{author}{\bibfnamefont{X.~G.} \bibnamefont{Wen}},
  \bibinfo{journal}{Phys. Rev. B} \textbf{\bibinfo{volume}{40}},
  \bibinfo{pages}{2719} (\bibinfo{year}{1989}).

\bibitem[{\citenamefont{Haerter and Shastry}(2005)}]{Haerter_Shastry}
\bibinfo{author}{\bibfnamefont{J.~O.} \bibnamefont{Haerter}} \bibnamefont{and}
  \bibinfo{author}{\bibfnamefont{B.~S.} \bibnamefont{Shastry}},
  \bibinfo{journal}{Phys. Rev. Lett.} \textbf{\bibinfo{volume}{95}},
  \bibinfo{pages}{087202} (\bibinfo{year}{2005}).

\bibitem[{\citenamefont{Kumar}(2008)}]{BK_new_rep}
\bibinfo{author}{\bibfnamefont{B.}~\bibnamefont{Kumar}},
  \bibinfo{journal}{Phys. Rev. B} \textbf{\bibinfo{volume}{77}},
  \bibinfo{pages}{205115} (\bibinfo{year}{2008}).

\bibitem[{\citenamefont{Lieb et~al.}(1961)\citenamefont{Lieb, Schultz, and
  Mattis}}]{XYchain}
\bibinfo{author}{\bibfnamefont{E.~H.} \bibnamefont{Lieb}},
  \bibinfo{author}{\bibfnamefont{T.~D.} \bibnamefont{Schultz}},
  \bibnamefont{and} \bibinfo{author}{\bibfnamefont{D.~C.}
  \bibnamefont{Mattis}}, \bibinfo{journal}{Annals of Phys. (N.Y.)}
  \textbf{\bibinfo{volume}{16}}, \bibinfo{pages}{407} (\bibinfo{year}{1961}).

\bibitem[{fno()}]{fnote_eg}
\bibinfo{note}{This expression for the ground state energy density in the
  thermodynamic limit matches with result of Shiba~\cite{Shiba}.}

\bibitem[{TVR()}]{TVR_comment}
\bibinfo{note}{This is in response to a question asked by Professor T. V.
  Ramakrishnan on the quantum decoherence temperature.}

\bibitem[{\citenamefont{Kugel and Khomskii}(1972)}]{Kugel_Khomskii}
\bibinfo{author}{\bibfnamefont{K.~I.} \bibnamefont{Kugel}} \bibnamefont{and}
  \bibinfo{author}{\bibfnamefont{D.~I.} \bibnamefont{Khomskii}},
  \bibinfo{journal}{Sov. Phys. JETP} \textbf{\bibinfo{volume}{15}},
  \bibinfo{pages}{629} (\bibinfo{year}{1972}).

\bibitem[{\citenamefont{Kumar and Majumdar}(2005)}]{Pinaki}
\bibinfo{author}{\bibfnamefont{S.}~\bibnamefont{Kumar}} \bibnamefont{and}
  \bibinfo{author}{\bibfnamefont{P.}~\bibnamefont{Majumdar}},
  \bibinfo{journal}{Eur. Phys. J. B} \textbf{\bibinfo{volume}{46}},
  \bibinfo{pages}{315} (\bibinfo{year}{2005}).

\bibitem[{\citenamefont{Shiba}(1972)}]{Shiba}
\bibinfo{author}{\bibfnamefont{H.}~\bibnamefont{Shiba}},
  \bibinfo{journal}{Phys. Rev. B} \textbf{\bibinfo{volume}{6}},
  \bibinfo{pages}{930} (\bibinfo{year}{1972}).

\end{thebibliography}
\end{document}